\documentclass[1p]{elsarticle}
\usepackage{amsmath}
\usepackage{braket}
\usepackage{csquotes}
\usepackage{lineno}
\title{Slotted Rotatable Target Assembly and Systematic Error Analysis for a Search for Long Range Spin Dependent Interactions From Exotic Vector Boson Exchange Using Neutron Spin Rotation}%\tnoteref{t1,t2}}
\author[nu]{C.~Haddock}%\corref{cor1}}%\fnref{fn1}}
\author[gc]{B. Crawford}%\fnref{fn2}}
\author[iub]{W. Fox}%\fnref{fn2}}
\author[ian]{{I. Francis}\fnref{fn1}}
\author[tt]{A. Holley}%\fnref{fn2}}
\author[gc]{S. Magers}%\fnref{fn2}}
\author[gtu]{M. Sarsour}%\corref{cor2}\fnref{fn1,fn3}}

\author[iub]{W. M. Snow}%\fnref{fn2}}
\author[iub]{J. Vanderwerp}%\fnref{fn2}}
\fntext[fn1]{This is the author's personal address.}
\address[iub]{ Indiana University/CEEM, 2401 Milo B Sampson Lane, Bloomington, IN 47408, USA}
\address[gtu]{ Georgia State University, 29 Peachtree
Center Avenue, Atlanta, GA 30303, USA}
\address[nu]{Nagoya University, Furocho, Chikusa Ward,
Nagoya, Aichi Prefecture 464-0814, Japan}
\address[gc]{Gettysburg College, 300 N Washington St,
Gettysburg, PA 17325, USA}
\address[tt]{Tennessee Tech University, 1 William L Jones
Dr, Cookeville, TN 38505, USA}
\address[ian]{612 S Mitchell St Bloomington, Indiana 47401, USA}
\begin{document}
\begin{abstract}
We discuss the design and construction of a novel target
array of nonmagnetic test masses used in a neutron
polarimetry measurement made in search for new possible
exotic spin dependent neutron-atom interactions of Nature at
sub-mm length scales. This target was designed to accept
and efficiently transmit a transversely polarized slow neutron
beam through a series of long open parallel slots
bounded by flat rectangular plates. These openings
possessed equal atom density gradients normal to the slots
from the flat test masses with dimensions optimized to
achieve maximum sensitivity to an exotic spin-dependent
interaction from vector boson exchanges with ranges in the
mm\,-\,$\mu$m regime. The parallel slots were oriented
differently in four quadrants that can be rotated about the
neutron beam axis in discrete $90^\circ$ increments using a Geneva drive. The spin rotation signals from the 4 quadrants were measured using a segmented neutron ion chamber to suppress
possible systematic errors from stray magnetic fields in the target region. We discuss the per-neutron sensitivity of the target to the exotic interaction, the design constraints, the potential sources of systematic errors which could be present in this design, and our estimate of the achievable sensitivity using this method.
\end{abstract}

\maketitle

\noindent Over the last decade a growing number of experiments have sought new interactions of Nature with weak couplings and
force ranges at the mm\,-\,$\mu$m scale. Such exotic
interactions might arise from string theory, from pseudo\,-\,Goldstone bosons generated by spontaneous symmetry breaking at high energy scales, from the as\,-\,yet\,-\,unknown physics behind dark energy, etc. A detailed review on the state of this developing subfield can be found in~\cite{Antoniadis11, Safronova2017}.\\

\noindent A general classification of the potentials that can exist between nonrelativistic fermions (protons, neutrons, and electrons in our case) from the nonrelativistic limit of a single spin 0 or spin 1 boson exchange assuming only rotational invariance~\cite{Dob06} uncovered 16 different operator structures at first order involving the spins, momenta, interaction range, and various possible couplings of the particles. This is a small enough number of distinct possibilities that it has motivated many
experimentalists to design specific experiments to look for
each type. The simple Yukawa interaction of range
$\lambda_c$ is the only spin\,-\,independent one on the list. The
rest contain Yukawa terms which set the distance scale for the interaction but also depend on the spins of one or both of
the fermions. 
\iffalse The mass $m_0$ of the exchange boson
is related to the range $\lambda$ of the Yukawa component
of the potential through the usual relation $m_0 =
\frac{\hbar}{\lambda\,c}$, where $\hbar$ is the reduced
Planck constant and $c$ is the speed of light in vacuum.\fi \\

\noindent Low energy neutrons are a particularly useful probe for new possible spin dependent interactions at the mm\,-\,$\mu$m scale, which could be mediated by a spin\,-\,1 boson whose
mass $m_0$ is related to this distance scale $\lambda_c$
through the relation $m_0 = \frac{\hbar}{\lambda_c\,c}$, where
$\hbar$ is the reduced Planck constant and $c$ is the speed
of light in vacuum. Slow neutrons can be polarized with high
efficiency and manipulated with delicate precision to conduct
sensitive interferometric measurements of many types. Such
measurements can be used to place limits on the strength of
new possible exotic couplings between ordinary matter.\\
 
\noindent We sought to measure the possible neutron-atom axial vector coupling given below in (\ref{eq:v5}): 
\vspace{0.25cm}
\begin{equation}
V_5=\frac{g_A^2\hbar^2}{4\pi m_n}\frac{e^{-
r/\lambda}}{r}\left(
\frac{1}{r}+\frac{1}{\lambda_c}
\right)
\vec{\sigma}\cdot\left(\frac{\vec{v}}{c}\times\frac{\vec{r}}{r}\right)\,.
\label{eq:v5}
\end{equation}
\vspace{0.25cm}

\noindent In this expression $g_A$ is the axial coupling constant, $r$
is the distance between the neutron and the atom, $v$ is the
relative velocity, $\sigma$ is the neutron spin, and $m_n$ is
the neutron mass. This interaction potential induces a
rotation of the spin about the $\vec{v}\times\vec{r}$ axis in a
way similar to a magnetic field. One can express the
effect of this potential in terms of a pseudomagnetic field, which if
integrated over a semi-infinite plane is given by
%~\cite{Pie11}
\vspace{0.25cm}
\begin{equation}
B_{AA}=\frac{1}{\gamma_n} \frac{g^2_A}{4} N \frac{\hbar
c}{m_nc^2} \lambda_c (\vec{v} \times \hat{y} ) e^{-\Delta y /
\lambda} ,
\label{eq:BAA}
\end{equation}
\vspace{0.25cm}

\noindent where $\lambda_c$ is the range of the force and the $y$
direction is normal to the face of the semi-infinite slab.\\

\noindent The first attempt to search for this exotic neutron-atom axial
vector coupling at interaction length scales below $1\,$cm
was made by Florian M. Piegsa and Guillaume Pignol at the Paul Scherrer Institute in Switzerland\,\cite{Pie11}. In their experiment the phase shift from this exotic interaction was sought by comparing the neutron phase shift difference from two parallel subbeams at a series of different distances from the same source (a copper plate in their case) and with the source placed close to one subbeam, with the other subbeam serving as a reference. The measurement used the well\,-\,known Ramsey
method of separated oscillatory fields and was therefore a
quantum interference experiment with a phase shift
proportional to the integral of the exotic interaction energy
between the neutron and the atoms in the slab. In neither subbeam do the neutrons touch the test mass surface. This approach renders the measurement insensitive to possible sources of systematic error from magnetic field drifts, thermal drifts which change the apparatus geometry, and the magnetic susceptibility of the test mass and has the advantage that it can fit for the known distance dependence of the exotic interaction. Possible systematic errors from magnetic impurities in the sample can be dealt with by measurements of the residual magnetic field from the sample. The ~sub-mm distance scales of interest for the exotic interaction search of interest dictates the choice of various aspects of the geometry.\\

%Since two beams are measured in parallel the relative phase shift can be deduced simultaneously and is therefore insensitive to effects from common mode magnetic field noise.\\

\noindent The number of neutrons used in the pioneering Piegsa and Pignol experiment was small and limited by their choice of a single planar source mass. We sought to improve upon this experiment by  increasing the total number of neutrons used to probe the possible spin dependent interaction and to employ polarized neutron spin rotation as the measurement method rather than Ramsey spectroscopy. The brightness of the slow neutron sources available for scientific research at national neutron facilities are all about the same within an order of magnitude or so as that of the PSI source used in their experiment, but many of the beams possess a much larger cross sectional area (as large as $10\,$cm $\times 10\,$cm) than that used by Piegsa and Pignol. However, we cannot take advantage of this full intensity as the potential we are trying to measure has a range of sub-mm scale. Since increasing the width of the beam necessarily increases the average distance between the neutron and slab, doing so quickly reduces the sensitivity to $V_5$ in the sub-mm range where it is poorly constrained experimentally.\\ 

\noindent Therefore rather than employing a single slab of material as the $V_5$ source, we designed a source target which consisted of an array of target slabs designed to maximize a possible $V_5$ signal. Our multi-slotted target design is an attempt to make efficient use of the large cross sectional area neutron beams which are available and still be able to access possible exotic effects of an interaction with sub-mm range. This mismatch of length scales immediately suggests a parallel multi-slotted target design. As this choice precludes the ability to vary the relative distance between source and subbeams as done by Piegsa and Pignol, we instead choose to surround each neutron subbeam through the target with masses of different densities (glass and copper in this case) and rotate the target so that the sign of the neutron phase shift from the exotic interaction is reversed. This allows us to conduct the search for this exotic interaction in the form of an asymmetry measurement by rotating the target mechanically. In place of the reference beam used by Piegsa and Pignol which is ideally far enough away from the source to feel no exotic interaction, we instead use a set of slots in a different quadrant of the target which are rotated by $90^\circ$ with respect to the neutron polarization direction. As can be seen from the form of the interaction, this different orientation gives a zero contribution to the exotic interaction after one averages over the sources. For one of the 4 orientations of the target one therefore gets a null phase shift in two diagonal quadrants and a positive and negative phase shift in the other two diagonal quadrants. One $90^\circ$ rotation of the target with respect to the beam reverses the positions of the low and high density sources with respect to the neutrons so that, in the absence of systematics, the phase shifts in the zero diagonal quadrants are still zero while the phase shifts in the other pair of diagonal quadrants are reversed. Four $90^\circ$ rotations of the target cycle all of the test masses through all quadrants of the beam to ensure that there are no effects from target nonuniformities.\\

\noindent In the reminder of this paper we briefly discuss the neutron spin rotation measurement method, the details of the target design used in the experiment, the potential sources of systematic error inherent in this method, and the prospects for improving on the sensitivity of the $V_{5}$ search.

\section{Neutron Spin Rotation Method}

\begin{figure}[!h]
\centering
\includegraphics[width=\textwidth]{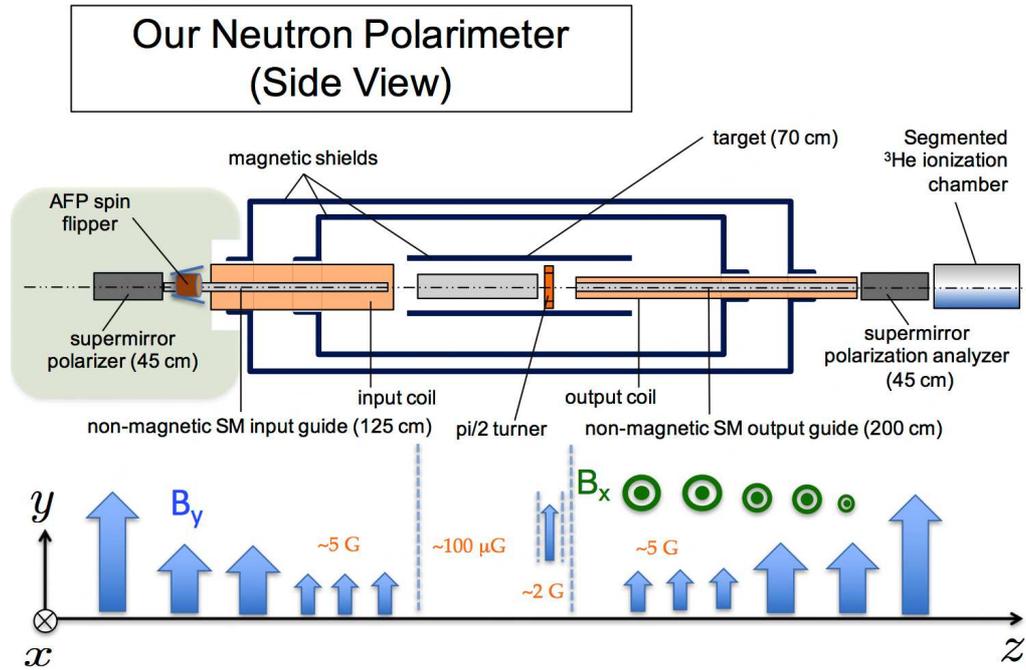}
\caption{Diagram of the Polarimeter used to measure the
longitudinal polarization asymmetry of the neutrons. This polarimeter is a slightly modified version of that described in \cite{NSR_pol}, where the target in that case is liquid helium and a $\pi$\,-\,rotation region exists in the center of the target region.}
\label{fig:polarimeter}
\end{figure}

\noindent We used a neutron spin rotation polarimeter to search for this interaction, employing a neutron polarimeter described in detail in~\cite{NSR_pol} and shown in Figure~\ref{fig:polarimeter}. This polarimeter is designed to accept a vertically polarized slow neutron beam and measure the horizontal component of the polarization that can result from a net rotation of the plane of polarization of the neutrons along the neutron beam axis in the magnetically-shielded sample region. This component of the neutron polarization is isolated by a combination of a precession of the neutron spin about a vertical axis in the sample region using an appropriately-tuned magnetic field combined with adiabatic spin transport in a magnetic field of oscillating helicity between the sample region and the neutron polarization analyzer. An earlier version of this polarimeter was used to search for parity violation in neutron spin rotation in $^{4}$He~\cite{Sno11}. The null result from this experiment was later used to constrain possible exotic parity-odd interactions of the neutron~\cite{Yan2013} and polarized neutron couplings to in-matter gravitational torsion~\cite{Lehnert2015} and in-matter nonmetricity~\cite{Lehnert2017}.\\

\begin{figure}[!h]
\centering
\includegraphics[width=\textwidth]{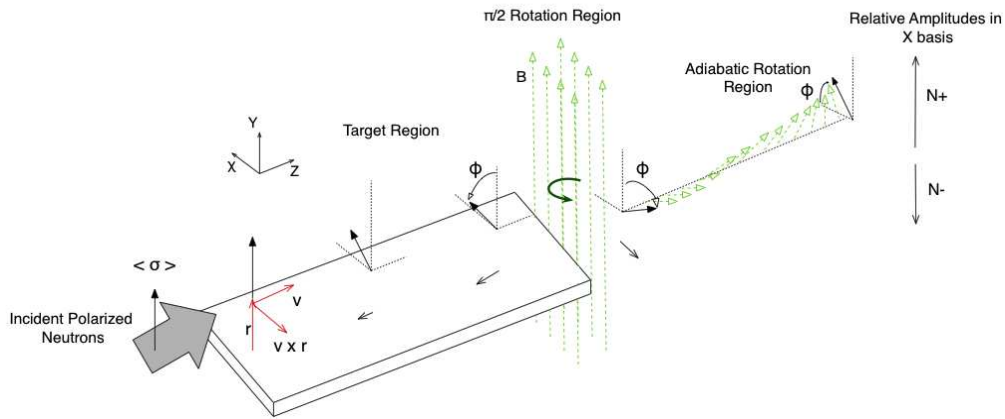}
\caption{The rotated spin expectation value is captured and
analyzed in a two step process. The \enquote{shadow} of the spin vector represents the net polarization asymmetry along a given axis.}
\label{fig:F5draw1}
\end{figure}

\noindent A slightly modified procedure described below was implemented to search for the neutron spin state change which would be caused by $V_{5}$. When a beam of polarized neutrons sent parallel to the
surface of a flat slab of nonmagnetic material is subject to
the effect of the potential in (\ref{eq:v5}), a rotation of the
expectation value of the neutron spin about an axis
perpendicular to both the spin and velocity is induced. As mentioned above, this can be understood from nonrelativistic quantum mechanics simply by viewing the cross product term in \ref{eq:v5} as a pseudomagnetic field about which the spin expectation value would Larmour precess. In order to measure the asymmetry in the
longitudinal polarization state of the neutron spin after it
passes through near a target slab we employ a two step
process (see Figure~\ref{fig:F5draw1}).
Once the neutron has accumulated an asymmetry from $V_{5}$ we rotate the polarization state by $\pi/2$ radians about the vertical
($+y$) axis using a constant magnetic field. We then rotate
this new polarization state by $ \pm \pi/2$ radians about the
longitudinal axis using the adiabatic neutron spin transport
field before finally arriving at a polarization analyzer which
allows through only neutrons whose polarization vector is
pointing along $+\hat{y}$ (see Figure~\ref{fig:polarimeter}).
By computing the difference ratio of the number of neutrons
that make it through the analyzer in the $+$ or $-$ state of
the adiabatic transport field we can measure an asymmetry which is proportional to the phase shift from the spin coupling of the neutron to the new possible light vector boson of interest.\\

\section{Alternating Density Gradient Scheme}
\noindent In order to increase the total number of neutron\,-\,atom
interactions over a polarized slow neutron beam of large
cross sectional area while remaining sensitive in the
mesoscopic length region of scientific interest, we designed
a target using multiple plates containing a large mass
density gradient. The test masses must be composed
of two plates of different densities. Plates of the same density would create an exact cancellation of any exotic spin\,-\,dependent interaction
coupling to atom density at the center and a significant
reduction near the edges due to the position vector\,-\,dependence of the sign of $V_5$. By using test masses with a large difference in mass density we ensure that the
neutron would see a net non\,-\,zero $V_5$ after summing
effects from all neutron trajectories between the plates.\\\\

\begin{figure}[h]
\centering
\includegraphics[width=\textwidth]{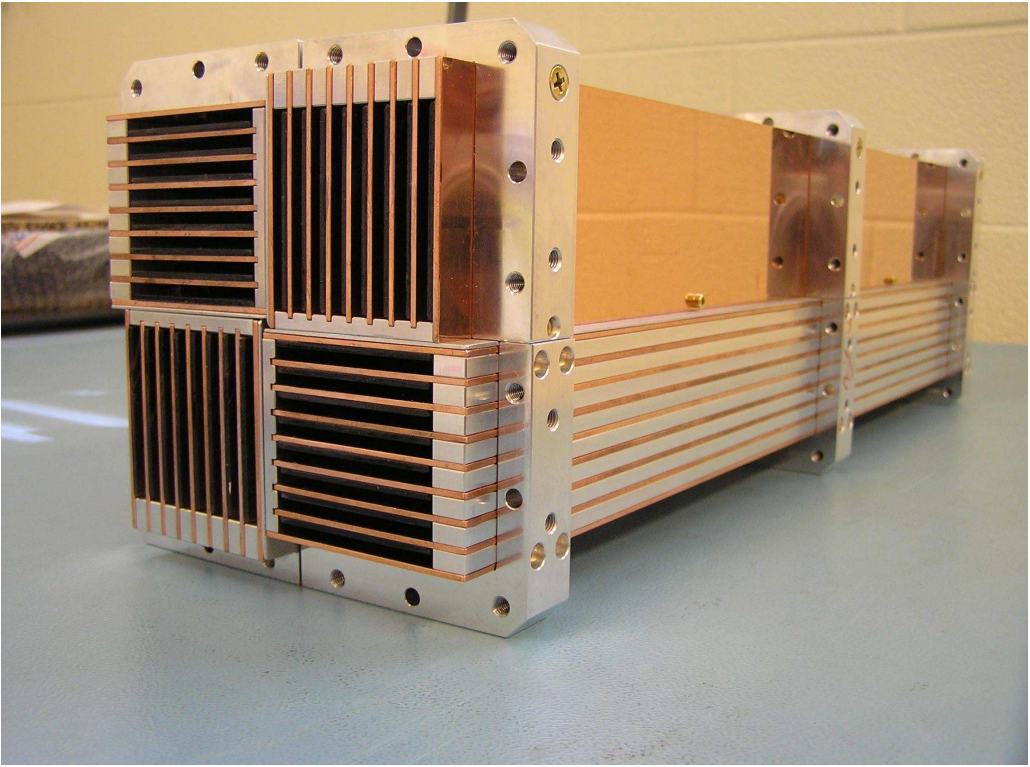}
\caption{The copper and float glass target array designed to
search for an exotic axial coupling of polarized neutrons to
matter. The borated aluminum neutron shielding used to reduce neutron activation of the target materials is not shown.}
\label{fig:target_bare}
\end{figure}

\noindent Neutron transport simulations were used to investigate the
sensitivity of the apparatus to $V_5$. A
simplified geometry of the target was inserted into an
existing neutron transport code written for previous
experiments with this apparatus studying parity violation in a
liquid helium target. The code uses realistic neutron
wavelength and angular distributions as well as neutron optical
transport. Using (\ref{eq:BAA}), the code calculated rotations for all trajectories,
including multiple bounces from the target faces, to estimate
the strength of the rotation signal and the statistical
precision. By varying target materials, thicknesses, neutron index of
refraction, and the size of the gap between target faces, a
set of viable materials and appropriate configurations was
obtained for the phase space of the slow neutron beam at
LANSCE, which is similar to that available on most fundamental neutron physics beamlines.\\

\noindent  One of the target plates was copper while
the other was float glass.
 The final choice of copper and float glass test masses was based on cost and availability, the low surface roughness of commercially-available materials, the reasonably-sized neutron optical potential of these materials, and on their well-known good neutron optical properties
(float glass has been used as backing for neutron supermirrors for
decades, and copper was one of the first materials used successfully in early neutron guides). Many of the neutrons passing through the target will bounce from the surfaces  of these materials since slow neutrons bounce off sufficiently flat surfaces, therefore, increase the neutron transmission through the device.\\ 

%Many of the neutrons passing through the
%target will bounce from the surfaces of these materials, and
%slow neutrons will bounce off sufficiently flat surfaces and
%therefore increase the neutron transmission through the
%device.
\noindent The neutron polarimeter used in this work contains a
neutron ion chamber at the end of the apparatus which is
split into four quadrants and can separately measure the
neutron spin rotation angle from each quadrant. The final
test mass slabs which fill the target are $50\,$cm long,
$6.5\,$cm wide and $1.65\,$mm thick. They are arranged in four quadrant regions each containing
eight neutron paths separated by two plate thicknesses. The
copper and float glass plates fit into $25\,\mu$m tolerance
parallel grooves that extend through the target region, as
shown in Figure~\ref{fig:target_bare}. As a control on the size
of rotations due to magnetic fields, two of the quadrants
have plates oriented such that $\vec{v}\times \vec{r}$ is in
the same direction as the initial neutron spin, in our case
vertical. Thus, one expects no fifth force rotation in the
quadrants with vertical plates.\newline

\noindent We can compare the sensitivity of this neutron beam/target combination with that of the PSI
experiment using the expression for the rotation resulting from a neutron
passing through the pseudomagnetic field near a slab of material. Using (\ref{eq:BAA}) we
can find the sensitivity to rotation as a function of $g^2_{A}$,
\vspace{0.25cm}
\begin{equation}
\frac{\phi_{PSI}}{g^2_{A}}=N L \frac{\hbar
c}{4 m_nc^2} \lambda_c e^{-\Delta y /\lambda} ,
\label{eq:phi_PSI}
\end{equation}

\noindent where $L$ is the length along which the neutron is in the pseudofield and N is the number density of the target slab. In the PSI experiment a pencil-like beam of approximately 0.3~mm$^2$ passed beside a 48-cm long, 1.9-cm thick copper slab. For the length scales of interest here the semi-inifinite slab approximation is sufficient. For $\lambda_c=\Delta y =0.1$~cm, we find $\phi_{PSI}/g^2_{A} = 5.2 \times 10^{10}$~rad. \newline

\begin{figure}[!h]
\centering
\includegraphics[scale=0.5]{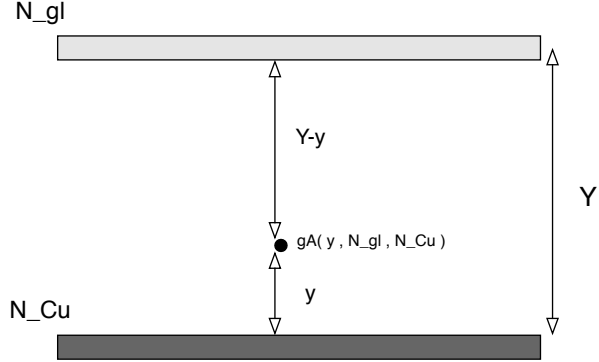}
\caption{Procedure for estimating the sensitivity in the air gap between two adjacent target plates using the new target design.}
\label{fig:approx}
\end{figure}

\noindent For the target described above we can make a similar estimate where neutrons are rotated in opposite directions from different density slabs above and below as shown in Fig.~\ref{fig:approx}. Averaging rotations along the vertical direction, $\frac{1}{Y} \int_0^Y dy$, between semi-infinite slabs separated by distance $Y$, we obtain
 
\begin{equation}
\frac{\phi_{NSR}}{g^2_{A}}= L \frac{\hbar
c}{4 m_nc^2} \lambda_c \frac{1}{Y} \int^Y_0 \left[ N_1 e^{-y /\lambda} - N_2 e^{-y /\lambda} \right] dy,
\label{eq:phi_NSR1}
\end{equation}
\vspace{0.25cm}
\noindent leading to
\vspace{0.25cm}
\begin{equation}
\frac{\phi_{NSR}}{g^2_{A}}= \frac{ L \lambda_c^2}{Y} \frac{\hbar
c}{4 m_nc^2} \left( N_1 - N_2 \right) \left(1 - e^{-Y /\lambda_c} \right).
\label{eq:phi_NSR2}
\end{equation}

\noindent Using the values of $L=50\,$cm and $Y=0.37\,$cm, the densities for copper and glass, and
$\lambda_c=0.1\,$cm, we find $\phi_{NSR}/g^2_{A} = 2.6 \times 10^{10}\,$rad, smaller than that obtained using (\ref{eq:phi_PSI}) by a factor of two. This approximation is less accurate at longer length scales as the targets no longer appear 
infinitely thick. To obtain a more realistic value for $\phi_{NSR}/g^2_{A}$ in our more complicated geometry we
performed Monte Carlo neutron transport simulations which use numerical solutions of the pseudofield between non-infinite slabs, include all slabs of the current target design, and incorporate the neutron phase space for FP12 at LANSCE. The simulations indicate about the same sensitivity as in the analytical calculation at $\lambda_c=0.01\,$cm, about a factor
of three less sensitivity at $\lambda_c=0.1\,$cm and a factor of 8 less sensitivity at $\lambda_c=0.3\,$cm. \\ 

\noindent For the $V_{5}$ length scales of interest in our work our target design has about an order of magnitude less sensitivity per neutron than the PSI experiment depending on the $V_{5}$ interaction range of interest. However this is more than compensated by the much larger beam size that this target design enables, which gives an improved statistical sensitivity relative to that experiment by at least a factor of $80$, which is the square root of the ratio of the cross sectional areas of the beams transmitted through the target region in both cases. Furthermore this does not take into account additional potential gains in sensitivity from the possible increased neutron density in the beam compared to the PSI experiment.\\ 

\noindent This design can therefore be used to perform a more sensitive search for $V_{5}$ on a large area cold neutron beam. However the additional types of systematic errors which this design can be susceptible to must be analyzed and properly suppressed to take full advantage of the greater potential statistical accuracy. We will discuss the potential sources of systematic  error below after giving a detailed description of the target design and operation.   

\iffalse
A final comparison: rotation angles seen in the statistics limited PSI experiment were in the $10^{-2}\,$rad range, while the previous NSR apparatus has been shown to access rotations in the $10^{-6}\,$rad range~\cite{NSR_pol} with the expectation that the new apparatus can reach $10^{-7}\,$rad.
\fi

\section{Target Rotation and State Detection}
\begin{figure}[!h]
\centering
\includegraphics[width=\textwidth]{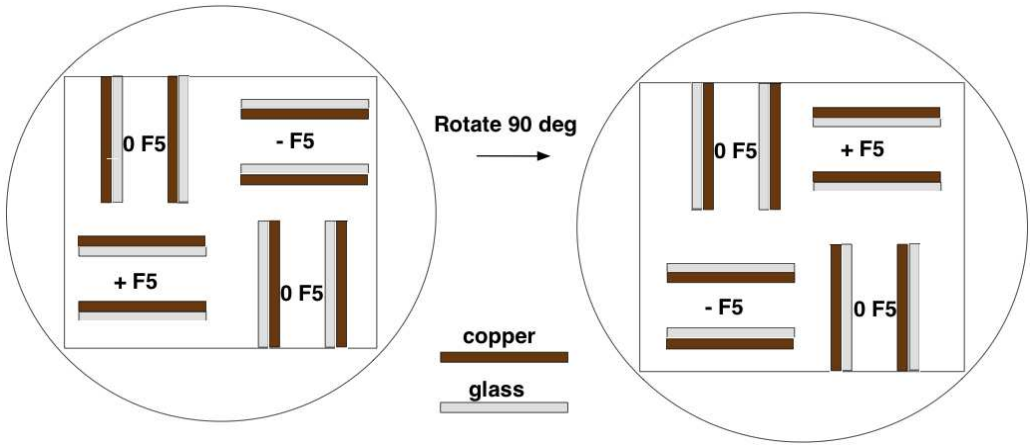}
\caption{A face\,-\,on simplified schematic of our target design. Neutrons pass through the region between two pairs of copper (brown) and glass (grey) target masses. Our alternating density gradient scheme reverses
the sign of the fifth force in each quadrant by rotating the target about its
longitudinal axis by $90^\circ$. Only 8 of the 32 pairs of test masses are depicted here for clarity.}
\label{fig:rot_drawing}
\end{figure}

\noindent To reduce the effect of possible space\,-\,dependent non\,-\,uniformities in the background magnetic field as well as possible differences in target plate properties (flatness,
thickness, etc.) it was crucial to have a mechanism to rotate
the target in discrete, repeatable $90^\circ$ increments. This
would allow neutrons to sample the same region of space
with different plates in the same orientation so that an
average can be carried out. Additionally by reversing the
direction of the mass gradient from quadrant to quadrant we
reverse the sign of $V_5$ between the 4 target states in
those quadrants for which $V_5$ is non zero, as depicted in
Figure~\ref{fig:rot_drawing}. Thus, by combining
measurements either from different target states or from diagonally
opposite quadrants, we can cancel rotations due to magnetic
fields while isolating the fifth-force rotations.\\

\noindent In order to consistently reproduce the four target states
between rotations about the longitudinal axis we utilized a so\,-\,called Geneva Drive mechanism which translates
continuous rotation into an intermittent rotary motion like in a
mechanical clock. This is done through the use of a rotating
cam with a pin which engages a slotted wheel attached to
the object to be rotated. The rotation of the object stops as
soon as the pin disengages while the cam may continue to
rotate independently before engaging another slot, providing
the discrete rotation mechanism desired. The rotating cam is driven by an air motor located outside of the outermost magnetic shielding to reduce the amount of magnetic components near the target. Standard lab air at $3\times 10^5\,$Pa pressure suffices to rotate the target. The flow to the air motor is controlled by the data acquisition system (DAQ) via an analog relay actuated valve. Each cam cycle rotates the target by $45^{\circ}$ and therefore two cycles were required
per target state rotation. Target state rotations took $2\,$s to
complete.\\

\begin{figure}[!h]
\centering
\includegraphics[width=\textwidth]{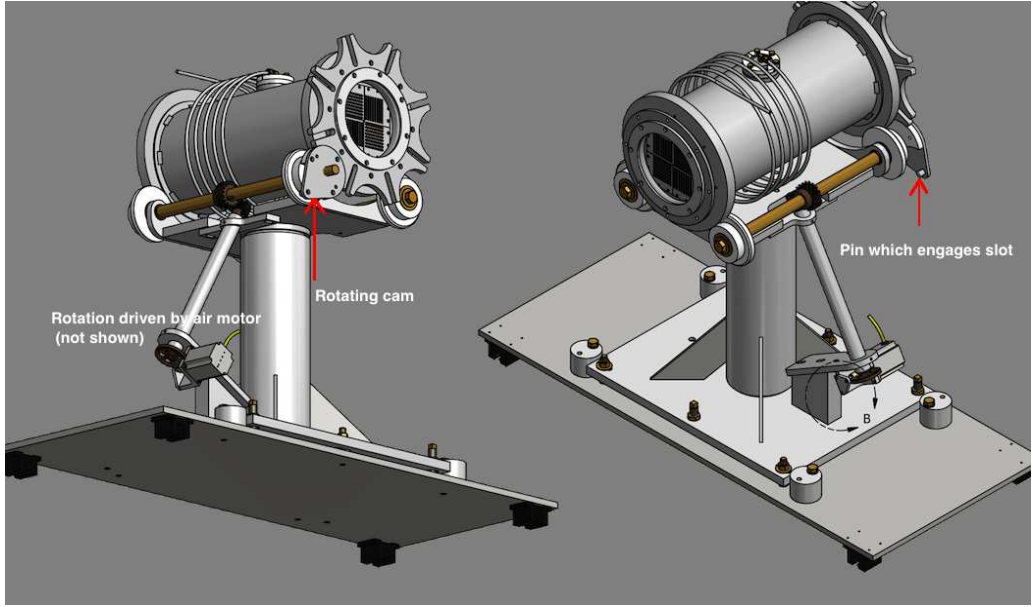}
\caption{CAD drawing of Geneva Drive used to rotate target
in discrete 45 degree steps.}
\end{figure}

%\newpage
\noindent To prevent the cam from over or under rotating, it is
crucial that we interrupt rotation when the target reaches
the desired state. Once the target spends adequate time in the
state, the cam will then begin rotation in the same
direction as previously to produce the next target state. The
target rotation will stop in the next target state using
an optical flag interruption scheme. We place an array of
three slotted $3\,$mm transmissive optical sensors below the
downstream face of the target such that the slots are
parallel to the target face. We fix four plastic optical flags
to the target face $90^\circ$ apart, each with either one,
two, or no holes. When the target is not in one of the four
desired states infrared light is transmitted to all three
receivers from their respective senders and each outputs a
non-zero voltage which is sent to the DAQ. The central sensor is always blocked for all four target
states and is unblocked during rotation. During the initial
fractions of a second when rotation starts, the DAQ ignores
the states of the flags while the flags move out of the target
position. Then once light transmission to the central sensor
is interrupted by the edge of the flag the DAQ waits $0.5\,$s
before stopping rotation. This wait time is long enough to
allow the cam to disengage the slotted wheel on the target
face but short enough to prevent reengagement from over
rotation. Once stopped, data is taken until it is time to rotate
to the next target state. Since each flag has a unique
arrangement of through holes, the target state can be identified by the DAQ. The DAQ code checks that the
registered state is the same as the intended state. If it is not,
the DAQ will continue to rotate the target until the intended
state is reached. This only occurrs at the
beginning of a run if the previous run is stopped before the
target reaches the usual starting state.\\

\section{Systematic Effects}

\noindent The systematic effects in a search for the $V_{5}$ interaction using neutron spin rotation with the target design presented in this paper possess some similarities to the types of systematic errors which have been considered in detail in our previous analysis~\cite{NSR_pol} of parity odd neutron spin rotation for polarized neutron transmission through matter. In both cases the main sources of systematic error come from the presence of residual magnetic fields in the target region coupled with some nonuniformity in the phase space of the neutron beam as it enters into and interacts with the target.\\

\noindent There are two main differences in the sources of systematic errors in these two types of measurements. Whereas for parity-odd neutron spin rotation one is mainly concerned with residual longitudinal magnetic fields, for a $V_{5}$ search one must consider systematic effects from both longitudinal and transverse residual magnetic fields in the target region.\\ 

\noindent In parity-odd neutron spin rotation the beam is transmitted through the target medium, and therefore target-induced nonuniformities in the neutron beam phase space may couple to residual magnetic fields and generate systematic effects. These effects tend to involve refractive neutron optics and small angle scattering, which bend the beam slightly so that it can sample a slightly different residual field in the presence of inevitable spatial gradients and yet still keep the transmitted beam within the phase space acceptance of the neutron polarization analyzer downstream. \\

\noindent In the case of a search for $V_{5}$ the beam is transmitted through a $^4He$ gas atmosphere in the target. Small angle scattering, which stays within the phase space acceptance of the polarimeter, combined with internal magnetic field gradients can generate a systematic error. This systematic error is very similar in character to the liquid helium small angle scattering systematic error described in great detail in \cite{NSR_pol} and is very small. This systematic is below $1\times10^{-7}\,$rad for our assumed internal magnetic field conditions.\\ 

\noindent We study other systematic effects by looking at the systematic errors which are common between the spin rotation approach to search for $V_{5}$ and the approach and implementation of the Piegsa and Pignol measurement:\\
\begin{enumerate}
\item The systematic error from possible magnetic impurities in the test masses is common to both approaches and can be bounded by performing magnetic measurements on the masses. In our target we searched for the presence of residual magnetic fields from the copper and glass plates using a fluxgate magnetometer. We saw no evidence for the presence of any such residual fields at the 10 $\mu$G level at a distance of $3\,$mm from the surfaces of the plates. The resulting upper bound on systematic errors in our measurement from this effect is below $2 \times 10^{-6}$ radians.\\ 
\item The systematic error from the magnetic susceptibility of the test masses, which shifts the value of any residual magnetic field as the mass is moved. This can be suppressed by making the residual magnetic field in the target region as small as possible. %In our case this was done with a three-layer mumetal magnetic shield which suppressed the residual magnetic field magnitude in the sample region to less than $1\,mG$. 
Using magnetic shielding it is possible to keep magnetic fields below $1\,$mG.
In our target this susceptibility systematic is  different in size from the PSI measurement due to two main effects:
\begin{itemize}
\item It is proportional to the difference in the magnetic susceptibilities of the two materials (copper and glass in our case) on either side of the slots, and 
\item It is slightly larger as some of the neutrons in our case bounce from the surface of the test masses and thereby encounter a slightly larger field difference than in the PSI measurement, where the neutron beam never touched the test mass. 
\end{itemize}

The magnetic susceptibility of both the copper and glass test masses in our target are of the same sign and possess magnitudes that are the same to about a factor of 3 which perturb the magnetic fields at the ppm level (the susceptibility of copper is about $\chi=-10\,$ppm and the susceptibility of glass is smaller). This leads to a systematic error in our apparatus of order $1 \times 10^{-4}\,$radians/Gauss if we make the extreme assumption that the neutrons move completely inside the matter. In fact the fraction of the target length that a  neutron is inside the copper during a reflection is much smaller than this, and this potential source of systematic error is utterly negligible.\\ 

\item Possible systematics from magnetic field drifts are suppressed in slightly different ways in the two designs as discussed above. Both conduct simultaneous null measurements with a reference beam. In the case of our rotating target, however, the simultaneous measurements are conducted on nominally identical but physically distinct test masses. The subsequent target rotations serve to test whether or not there are any nonuniformities in the test mass properties. One $90^\circ$ rotation of the target with respect to the beam reverses the positions of the low and high density sources with respect to the neutrons, thus removing non-target related rotations such as from stray magnetic fields. Therefore, in the absence of systematics, the phase shifts in the zero diagonal quadrants are still zero while the phase shifts in the other pair of diagonal quadrants are reversed.\\
\end{enumerate}

\noindent In the PSI experiment the neutron beam did not touch the copper test mass. However the beam bounces off the surface of the plates in our design and so we have to consider a different set of systematic errors associated with neutron reflection. We note two potential forms of systematic error which are present in our target design but not in the PSI experiment come from the neutrons bouncing off the test mass surfaces. One comes from the fact that, due to the different neutron indices of refraction of the two test masses, the sections of the neutron phase space transmitted by the target will be very slightly different in the different target positions. This can generate a potential systematic error if these slightly different neutron beam phase space sections see different magnetic fields. This source of systematic error can be suppressed by minimizing both the absolute B fields in the target region and also B field gradients and is discussed in more detail below.\\
 
\noindent Another potential source of systematic error comes from the very small changes in the neutron beam polarization from neutron spin-orbit scattering from the test masses, which contains the same operator structure as that from the $V_{5}$ interaction of interest. 
The effect on the polarization from neutron spin-orbit scattering is proportional to the neutron momentum transfer q, which, although nonzero for neutron optical reflection from a mirror, is quite small. It is also small because the neutron spin-orbit scattering, which is classically a velocity-dependent effect, leads to an imaginary scattering amplitude, and as the neutron-nucleus scattering amplitude is mainly real in the absence of n-A resonances the neutron spin-dependent component of the interference of the spin-orbit scattering with the potential scattering is a quadratic effect. 
This latter effect can be calculated using the nice formulae in~\cite{Gericke2008} and is very small in our case: for copper the maximum analyzing power from polarized neutron/copper atom scattering which occurs for momentum transfers corresponding to the critical angle for total external reflection is less than $5 \times 10^{-9}$ and is therefore completely negligible. Although some neutrons can scatter at higher momentum transfers from the mirrors in the diffuse reflection component of the beam, the intensity of the beam in this component is small compared to the specular component and almost all of the diffusely-reflected neutrons fall well outside the phase space acceptance of the rest of the apparatus. Therefore in the other systematics effects discussed below we will concentrate only on those coming from specular reflection. \\

%\iffalse
\noindent To reduce the effects of common-mode magnetic field noise on the neutron spin rotation signal, we arranged the target masses such that different regions of the beam area were made sensitive to possible $V_5$ signals of opposite sign. By recording these rotation angles simultaneously using a segmented neutron detector we are able to remove the effect of common field noise by taking the difference of these rotation angles. To further reduce systematic effects arising from stray magnetic fields we designed the target to rotate about the beam axis which, due to the arrangement of test masses, changes the sign of $V_5$ in each region. This allows for a comparison of rotation angles in the same beam phase space at different times, thereby removing the effect of space-dependent background field gradients whereas the aforementioned simultaneous measurement of rotations of opposite sign would remove the effect of time-dependent background field fluctuations.\\ 

\noindent The spin state changes due to longitudinal or transverse magnetic fields naturally generate different sorts of systematic effects. Longitudinal fields rotate the polarization vector along the axis of the neutron momentum (\enquote{left/right} rotations). Transverse magnetic fields rotate the polarization vector forward/backward along the neutron momentum and therefore directly mimic the effect of $V_{5}$.  Rotations by longitudinal fields before the $\pi/2$ coil are not analyzed by the downstream polarization analyzer but they reduce the polarization product PA by a factor of $\cos{\theta_{B_{L}}}$. The $\pi/2$ coil turns forward/backward $V_5$ rotations into left/right rotations to be analyzed by the analyzing super mirror. Thus, left/right rotations from a longitudinal field after the $\pi/2$ coil add to or subtract from the desired signal and thus can cause false effects.  Rotations from transverse fields after the $\pi/2$ coil reduce the PA value by $\cos{\theta_{B_{T}}}$. \\

\noindent There are three main effects that can affect the size of a nonzero signal from $V_5$ in our setup. Nonzero rotations from fringing pseudomagnetic fields at the edges of vertical-plate quadrants can dilute the signal when subtracting horizontal and vertical quadrants.  Magnetic field gradients which cause slightly different transverse and longitudinal fields in each of the four quadrants can lead to residual rotations after subtracting rotations from different quadrants.  A third phenomenon which can both dilute the signal and also lead to a systematic effect comes from \enquote{cross-over} neutrons which pass through the target in one quadrant but appear downstream in a different quadrant due to beam divergence in the space between the $\pi/2$ coil and the entrance of the output guide as well as beam transport through the guide (which is split into two separate guide sections with a vertical septum with a supermirror coating on both sides). Rotations from transverse and longitudinal fields from the other three quadrants therefore get mixed into the total rotation of each quadrant.  A clean cancellation of magnetic field systematics by doing diagonal averaging/subtraction therefore depends on the amount and symmetry of the signals from these cross-over neutrons as well as the effect of magnetic field gradients.\\

\noindent To analyze the effects of these systematics we present the procedure used to extract the $V_{5}$ signal. Let $Q1$, $Q2$, $Q3$, and $Q4$ be the signals in the 4 quadrants of the ion chamber, where we assume a quadrant arrangement with target sections 1 and 3 possessing horizontal target plates, target sectors 2 and 4 possessing vertical target plates, and the four quadrants starting with 1 in the upper right corner proceed numerically in a counterclockwise direction (see Figure \ref{fig:rot_drawing}).   The rotation angle from $V_5$ in the horizontal (vertical) target region is given by $\theta_{HF5}$ ($\theta_{VF5}$), where $\theta_{HF5}$ is the desired signal and $\theta_{VF5}$ is a rotation due to fringing of the pseudomagnetic field at the edges of vertical plates.  We let $\theta_{1BTU}$ ($\theta_{1BLD}$) be the integrated spin rotation angle from the average upstream transverse (downstream longitudinal) field of the first quadrant and similarly for the other three quadrants.  To incorporate the possibility of cross-over neutrons, we let $a_{1}$\,,\,$a_{2}$\,,\,$a_{3}$\,, and $a_{4}$ be the fractions of counts in the different ion chamber quadrants from side-to-side cross-over neutrons originating in the quadrant given by the associated subscripts and $b_{1}$\,,\,$b_{2}$\,,\,$b_{3}$, and $b_{4}$ be the fraction of counts from up-down cross-over neutrons originating in the quadrant given by the associated subscript. Note that quadrants 1 and 4 have opposite $\pi/2$ rotations from quadrants 2 and 3 so $\theta_{xBTU}$ changes sign depending on the quadrant. Target quadrants 1 and 3 have opposite target orientations as do target quadrants 2 and 4, so coupled with the $\pi/2$ left/right sign change $\theta_{HF5}$ and $\theta_{VF5}$ have the same left/right sign, as do $\theta_{xBLD}$ as they are downstream from the $\pi/2$ coil. We ignore possible diagonal cross-over neutrons which are negligible by design and in simulations.  All angles are assumed to be small enough that we can simply add rotations; this assumption is true for the residual field values we consider. In this case, we get

\begin{align}
Q1&=\theta_{HF5} + \theta_{1BTU}+\theta_{1BLD}+a_{2} \left(-\theta_{2BTU}+\theta_{2BLD}+\theta_{VF5} \right)\label{eq:magneticsystematics} \\& + b_{4} \left( \theta_{4BTU}+\theta_{4BLD}+\theta_{VF5} \right)   \nonumber \\
Q2&=\theta_{VF5} -\theta_{2BTU}+\theta_{2BLD}+ a_{1} \left(\theta_{1BTU}+\theta_{1BLD}+\theta_{HF5} \right) \nonumber\\& + b_{3} \left(-\theta_{3BTU}+\theta_{3BLD}+\theta_{HF5} \right)   \nonumber \\
Q3&=\theta_{HF5} -\theta_{3BTU}+\theta_{3BLD}+ a_{4} \left(\theta_{4BTU}+\theta_{4BLD}+\theta_{VF5} \right) \nonumber\\& + b_{2} \left(-\theta_{2BTU}+\theta_{2BLD}+\theta_{VF5} \right)   \nonumber \\
Q4&=\theta_{VF5} + \theta_{4BTU}+\theta_{4BLD}+a_{3} \left(-\theta_{3BTU}+\theta_{3BLD}+\theta_{HF5} \right) \nonumber\\& + b_{1} \left(\theta_{1BTU}+\theta_{1BLD}+\theta_{HF5} \right) .  \nonumber \\ \nonumber
\end{align} 

\noindent Now we assume that quadrant differences in the transverse and longitudinal fields are due to linear field gradients in the horizontal and vertical directions, such that the quadrant average of the rotation due to transverse (longitudinal) is given by $\theta_{BTU}$ ($\theta_{BLD}$) and is modified by $\pm \delta_H \theta_{BTU}$ ($\pm\delta_V \theta_{BTU}$) due to a horizontal (vertical) gradient for the transverse field, and similarly for the longitudinal field.  Thus, the following terms that arise in the next step where we add diagonal quadrants can be written as
\begin{align}
 \frac{1}{2}\left(\theta_{1BTU}-\theta_{3BTU}\right)  & = \delta_H \theta_{BTU} + \delta_V \theta_{BTU} \\
 \frac{1}{2}\left(-\theta_{2BTU}+\theta_{4BTU}\right) &= \delta_H \theta_{BTU} - \delta_V \theta_{BTU} \nonumber \\
 \frac{1}{2}\left(\theta_{1BLD}+\theta_{3BLD}\right) &=  \theta_{BLD}  \nonumber \\
 \frac{1}{2}\left(\theta_{2BLD}+\theta_{4BLD}\right) &=  \theta_{BLD} . \nonumber \\ \nonumber 
\end{align}
\noindent Averaging diagonal quadrants cancels rotations due to the average upstream transverse field and gradients in the longitudinal field:

\begin{align}
(Q1+Q3)/2 \label{eq:magneticsystematicdiagonalaverage2}\\ &=\theta_{HF5} + \left(\delta_H \theta_{BTU} + \delta_V \theta_{BTU}\right)  + \theta_{BLD} \nonumber  \\& 
   + 1/2 \left(a_{2}+b_{2}\right) \left(-\theta_{2BTU}+\theta_{2BLD}+\theta_{VF5} \right) 
   + 1/2 \left(a_{4}+b_{4}\right) \left(\theta_{4BTU}+\theta_{4BLD}+\theta_{VF5} \right) \nonumber \\
(Q2+Q4)/2 \nonumber\\ &=\theta_{VF5} + \left( \delta_H \theta_{BTU} - \delta_V \theta_{BTU} \right) + \theta_{BLD}  \nonumber  \\& 
   + 1/2 \left(a_{1}+b_{1}\right) \left(\theta_{1BTU}+\theta_{1BLD}+\theta_{HF5} \right) 
   + 1/2 \left(a_{3}+b_{3}\right) \left(-\theta_{3BTU}+\theta_{3BLD}+\theta_{HF5} \right) .\nonumber
\end{align}

\noindent Subtracting the two averages above (\ref{eq:magneticsystematicdiagonalaverage2}) for data taken at the same time further reduces common mode noise and eliminates rotations from downstream longitudinal fields and the effect of a transverse field horizontal gradient and leaves us with 

\begin{align}
\phi_{diag} = (Q1+&Q3)/2 - (Q2+Q4)/2  = \label{eq:magneticsystematicdiagonalaverage3}\\
 & \theta_{HF5} - \theta_{VF5} + 2\delta_V \theta_{BTU} \nonumber \\
 &- 1/2 \left(a_{1}+b_{1}+a_{3}+b_{3}\right)\theta_{HF5} + 1/2 \left(a_{2}+b_{2}+a_{4}+b_{4}\right) \theta_{VF5}  \nonumber \\
 &+ 1/2 \left(a_{2}+b_{2}\right) \left(-\theta_{2BTU}+\theta_{2BLD}\right) \nonumber \\
 &+ 1/2 \left(a_{4}+b_{4}\right) \left(\theta_{4BTU}+\theta_{4BLD} \right) \nonumber \\
 &- 1/2 \left(a_{1}+b_{1}\right) \left(\theta_{1BTU}+\theta_{1BLD} \right)  \nonumber \\
 &- 1/2 \left(a_{3}+b_{3}\right) \left(-\theta_{3BTU}+\theta_{3BLD} \right) \nonumber \\ \nonumber
\end{align}

\noindent Here we see the three effects mentioned earlier: the reduction in the $\theta_{HF5}$ signal by $\theta_{VF5}$ from fringing pseudomagetic fields in vertical targets, a remaining effect from vertical transverse field gradient, and a number of terms due to neutrons crossing between the four quadrants after the $\pi/2$ coil which modify the $V_5$ signals and introduce false effects from non-zero magnetic fields in the target region.  If there are no gradients in the magnetic fields we can simplify the above expression,

\begin{align}
\phi_{diag}^{noGrad} =(Q1+&Q3)/2 - (Q2+Q4)/2  = \label{eq:magneticsystematicdiagonalaverage4}\\
 & \theta_{HF5} - \theta_{VF5} \nonumber \\
 &- 1/2\left(a_{1}+b_{1}+a_{3}+b_{3}\right)\theta_{HF5} + 1/2 \left(a_{2}+b_{2}+a_{4}+b_{4}\right) \theta_{VF5}  \nonumber \\
 &+ 1/2 \left( \left(a_{3}+b_{3}+a_{4}+b_{4}\right)-\left(a_{1}+b_{1}+a_{2}+b_{2}\right) \right) \theta_{BTU}  \nonumber \\
 &+ 1/2 \left( \left(a_{2}+b_{2}+a_{4}+b_{4}\right)-\left(a_{1}+b_{1}+a_{3}+b_{3}\right) \right) \theta_{BLD}  \nonumber \\ \nonumber
\end{align}

\noindent The top and bottom pairs of quadrants both contain horizontal and vertical targets, so if everything is aligned well the total cross-over neutrons from bottom quadrants should equal the number from the top quadrants and then the contribution $\theta_{BTU}$ from residual transverse magnetic fields is small. ¨The systematic error from the longitudinal residual field rotation is a different story, since the combination of a larger divergence in the vertical direction and the reduction in cross-over neutrons from the vertical septum conspire to make the cross-over neutrons from the vertical targets much larger than the cross-over neutrons from the horizontal targets.  Since the longitudinal field is a problem only after the $\pi/2$ coil, this effect can be reduced by reducing the separation between the end of the $\pi/2$ coil and the start of the output coil.  For typical numbers, which one can expect in our apparatus with an aligned beam and a $2\,cm $ gap after the $\pi/2$ coil, one gets from simulation the following fractions for cross-over neutrons : $a_{1}=0.0148 $, $ b_{1}=0.0083$ , $a_{2}=0.0064 $, $b_{2}=0.1952$, $a_{3}=0.0145 $, $b_{3}=0.0084$, $a_{4}=0.0066 $, $b_{4}=0.1963 $,\,which results in $(Q1+Q3)/2 - (Q2+Q4)/2 = 0.9770\,\theta_{HF5} - 0.7978\,\theta_{VF5} + 0.0006\,\theta_{BTU} + 0.1793\,\theta_{BLD}$.  Simulations show a $<2 \times 10^{-7}$ radian systematic error in a $0.1\,mG$ field. If one adds in addition a $1\,mrad$ neutron beam misalignment then the cross-over neutron fraction from the bottom left ($Q3$) will be more than from the top right ($Q1$), and this difference between the bottom versus top quadrants generates a systematic error from the residual transverse magnetic field. For our assumed residual fields and geometry we get a $2 \times 10^{-6}$ radian systematic effect.  \\

\noindent Magnetic field gradients pose a potentially larger problem. Typical gradients lead to quadrant field differences of less than 10\%, so for $6\,\AA$ neutrons 
in a $0.1\,$mG field, $ 2 \delta_V \theta_{BTU} < 1.5 \times 10^{-4}\,$rad. 
The cross-over neutrons modify this systematic at less than the 10\% level.  Therefore, while the analysis so far reduces common mode noise, it is not sufficient to remove all magnetic-field related systematic effects to the desired level.  To further reduce systematic effects from magnetic fields, we rotate the target by $90^{\circ}$ thereby flipping the orientation of the density gradient and thus the signs of the $V_5$ effects while leaving rotations from magnetic fields unchanged.  By taking the difference in $\phi_{diag}$ (\ref{eq:magneticsystematicdiagonalaverage3}) before and after a $90^{\circ}$ rotation of the target, we eliminate magnetic field rotations while retaining the reduction in common-mode noise,

\begin{align}
\Phi_m = \frac{\left( \phi_{diag} - \phi_{diag}' \right)}{2} = 
 &\left[1-1/2 \left(a_{1}+b_{1}+a_{3}+b_{3}\right)\right] \theta_{HF5}  \label{eq:magneticsystematicdiagonalaverage5} \\ 
 & -  \left[1 - 1/2 \left(a_{2}+b_{2}+a_{4}+b_{4}\right)\right] \theta_{VF5}  .  \nonumber \\ \nonumber
\end{align}

\noindent  Using the cross-over fractions from simulation, we find $\Phi_m \approx 0.98 \,\theta_{HF5} - 0.80\,\theta_{VF5}$, and since $\theta_{VF5} < 0.1\,\theta_{HF5}$ we find about a 10\% reduction in the desired signal, $\Phi_m \approx 0.9\,\theta_{HF5}$.  However, as noted above the subtractions in this last step are done for rotations using different plates, so plate non-uniformities from quadrant to quadrant can generate a systematic error.  Consider a mechanical imperfection of the target which generates a $1\,$mrad angular twist in the plate orientations in quadrant 1 after a $90^\circ$ target rotation. Our simulations show a systematic effect of almost $1 \times 10^{-6}\,$rad for our assumed residual field values. This is only an issue if the $1\,$mrad polarization twist in the plates is in the direction opposite the target orientation, e.g. a $1\,$mrad vertical twist in horizontal target plates.\\

\noindent Another systematic effect in this case can come from differences in the reflectivity of the different pieces of copper, which might not be identical for the two $90^\circ$ states. Consider a very extreme case in which the copper plates sensitive to $V_{5}$ in quadrants/states $Q1$ and $Q3$ possess $100\%$ reflectivity and the different copper plates in the same location in quadrants/states $Q1$ and $Q3$  have completely non-reflective copper plates. In this case one gets a systematic error of $4 \times 10^{-6}\,$rad.  Even if such an extreme case were to be realized by some chance between each pair of plates there is no reason that they should be correlated in sign, so one would expect the systematic error from the difference between the average reflective properties of each twelve-plate quadrant to be suppressed by at the very least a factor of $\sqrt{12}$. One can easily perform visual inspection and measurements of the surface roughness of the plates along with neutron reflectometry measurements of the properties of the individual plates themselves to further constrain and suppress this potential source of systematic error.\\

\begin{table}[h]
\caption{A list of sources for potential systematic effects in a search for the $V_{5}$ interaction using a slow neutron polarimeter in combination with the target design presented in this paper. These estimates all hold for the internal longitudinal and transverse magnetic fields of $0.1\,$mG assumed. We have included all systematic errors associated with analysis after both modes of target cancellation (diagonal averaging followed by $90^\circ$ target rotation). All of the dominant sources of systematic error on this list scale with the size of these residual internal fields.}
\begin{center}
%\caption{}
\label{tbl:sys}
\begin{tabular}{lllclcl}
\hline
Source & Systematic (rad) \\
small angle scattering from $^4He$ atmosphere: & $< 1 \times 10^{-7}$ \\
target mass diamagnetism: & $1 \times 10^{-9}$ \\
neutron-atom spin-orbit scattering: & $5 \times10^{-9}$ \\
target magnetic impurities: & $<2\times10^{-6}$ \\
target misalignment: & $<1\times 10^{-6}$ \\
target reflectivity differences: & $<1\times10^{-6}$ \\
\hline
\end{tabular}
\end{center}
\end{table}

\noindent Table 1 shows our estimates for the sizes of the various forms of systematic error for our target design and measurement methods. Almost all of the systematics are associated with residual magnetic fields coupled with various types of apparatus or beam nonuniformities. The different subtractions enabled by our target design reduce both common-mode noise and systematic errors to the desired level.

\section{Conclusions}
\noindent The Neutron Spin Rotation collaboration developed a target
consisting of alternating plates of different mass materials for
use in experiments seeking new spin\,-\,dependent 
interactions using polarized slow neutrons. The four chambers of the
target allow cancellation of neutron spin rotations from stray
magnetic fields. A Geneva Drive rotates the
target such that the plate orientations flip in alternate target
states, thereby further canceling magnetic field rotations
while isolating rotations from new interactions. The target was
used in a recent experiment on the FP12 cold neutron beam
at LANSCE, the results of which will be presented in a
forthcoming paper.\\

\noindent The sensitivity of this target for this exotic interaction
search could be further improved by using a denser
nonmagnetic mass such as tungsten or tantalum in place of
the copper used in this experiment. The higher density would
lead to a greater sensitivity to this possible exotic interaction
from the larger number of electrons and nucleons per unit
volume in the source. In addition it would be desirable to
flatten and polish the surfaces of the test masses exposed to
the neutron beam well enough that slow neutrons are
guaranteed to undergo specular reflection from the
surfaces. This would raise the fraction of neutrons
transmitted through the apparatus and increase the
sensitivity of the experiment to shorter interaction ranges. The target can be made longer in principle as long as the magnetic shielding can be maintained.\\

\noindent We can estimate the order of magnitude of the sensitivity to the $V_{5}$ interaction that could be achieved in a dedicated experiment using a fully optimized target design. Simulations of the neutron spin rotation apparatus for a proposed measurement of parity odd neutron spin rotation in liquid helium (whose statistical error has already been thoroughly analyzed) adapted for this target show that given two months of dedicated beam time it is possible to increase the sensitivity in $g_A^2$ by 4 orders of magnitude compared to the pioneering work of Piegsa and Pignol. If one can lower the systematic errors further by suppressing the residual longitudinal and transverse magnetic fields in the target region below 100 $\mu$G, one could probe neutron axial couplings to matter through an exotic spin 1 boson which are about $13$ orders of magnitude weaker than electromagnetism.

\section{Acknowledgements}
%Since we would be accessing the cave to make
%adjustments throughout the run cycle we sought to reduce
%the activity levels near the target region as much as
%possible. This was done by installing 3 1mm-thick Borated
%Aluminum masks to the upstream end of the target face to
%reduce neutron absorption in the edge of the Cu plates.

\noindent We would like to thank Phil Childers and the staff of the
Indiana University Swain Hall physics machine shop for an
outstanding machining job. C. Haddock and W. M. Snow
acknowledge support from US National Science Foundation
grants PHY-1306942 and PHY-1614545 and by the Indiana
University Center for Spacetime Symmetries. C. Haddock
acknowledges support from the US Department of Energy
SCGSR Fellowship and the Japan Society for the Promotion
of Science Fellowship. M. Sarsour acknowledges support
from US Department of Energy grant DE-SC0010443.\\

%\section*{References}

\end{document}